\documentclass[epj]{svjour}
\usepackage{graphicx,amsmath,ifpdf}

\begin{document}

\title{Interplay among the azimuthally dependent HBT radii and the elliptic flow}
\subtitle{Insights from the Buda-Lund hydro model}
\author{M\'at\'e~Csan\'ad\inst{1}
   \and Boris Tom\'a\v{s}ik\inst{2,3}
   \and Tam\'as Cs\"org\H{o}\inst{4}}
\institute{
E\"otv\"os University, H- 1117 Budapest XI, P\'azm\'any P\'eter s. 1/A, Hungary
\and
Univerzita Mateja Bela, 97401 Bansk\'a Bystrica, Slovakia
\and
FNSPE, Czech Technical University, Prague, Czech Republic
\and
MTA KFKI RMKI, H-1525 Budapest 114, POBox 49, Hungary}
\date{\today}

\abstract{We present a calculation of the elliptic flow and azimuthal
dependence of the correlation radii in the ellipsoidally symmetric
generalization of the Buda-Lund model. The elliptic flow is shown to
depend only on the flow anisotropy while in case of correlation radii
both flow and space anisotropy play an important role in
determining their azimuthal oscillation. We also outline a simple
procedure for determining the parameters of the model from data.}


\maketitle
%%%%%%%%%%%%%%

\section{Introduction}

Studies of the freeze-out configuration in non-central
ultra-relativistic nuclear collisions have yielded valuable
information on the dynamics of the collision. The azimuthal asymmetry
of hadronic momentum spectra---widely known under the somewhat
vague term ``elliptic flow''---is a cornerstone result for
conclusions about thermalization and near perfect fluid dynamics~\cite{Riordan:2006df}.
The measurement of azimuthal dependence of correlation radii in
non-central collisions indicates a fireball which breaks up while
still being elongated out-of-reaction-plane~\cite{Retiere:2003kf}. This is the original
orientation of the system and the observation thus provides an
upper bound on the total lifespan of the created fireball and suggests an early
and sudden freeze-out in a blast-wave picture, influentially suggested in ref.
~\cite{Schnedermann:1993ws}.

These observables can result from an interplay of spatial asymmetry of the
fireball and the asymmetry of the flow velocity field of an exploding fireball.
The question is if and how these asymmetries can be both determined from data. Such  a
question can be studied in the framework of a model. Here we present a study that utilizes
the Buda-Lund hydro model.

For the present study we use the ellipsoidal Buda-Lund
parametrization of the freeze-out of the fireball
\cite{Csanad:2004cj}. Note that the parameterization corresponds to
a solution of certain class of hydrodynamic
 models~\cite{Csorgo:2001xm,Csorgo:2003rt,Sinyukov:2004am,Csorgo:2006ax}.

We shall show that in the Buda-Lund model elliptic flow results solely from
the flow asymmetry, and that the entanglement of spatial and flow anisotropy in
determining the azimuthal oscillation of the correlation radii is rather strong.
This can be contrasted to other derivations that explore the conditions
for quark coalescence~\cite{Pratt:2004zq}, or the numerical solutions of
boost-invariant, azimuthally dependent hydrodynamics with Gaussian
initial conditions ~\cite{Broniowski:2008vp}. Azimuthally asymmetric HBT radii
were also considered recently in cascade models, e.g. in the fast Monte-Carlo
model of ref.~\cite{Amelin:2007ic}, or, in the Hadronic Resonance Cascade~\cite{Humanic:2005ye}. A simple procedure to disentangle spatial and momentum space anisotropy
from elliptic flow and azimuthally sensitive HBT measurements was proposed
for generalized blast-wave type of models in ref.~\cite{Tomasik:2005ny}.
Now we investigate how to disentangle these effects in the ellipsoidally
symmetric version of the Buda-Lund hydro model.
Let us note that we cannot attempt to evaluate or fully cite all the relevant
works in this technical paper, but we kindly refer the interested readers to the
review of reviews in ref.~\cite{Csorgo:2005gd}, and also to ref.~\cite{Lisa:2005dd},
a recent review that emphasizes the femtoscopy aspects of particle interferometry.

The structure of our paper is as follows:
In Section~\ref{model} we shall shortly introduce the basic features of the
ellipsoidally symmetric Buda-Lund model.
Section~\ref{vtwo} deals with the calculation of elliptic flow and Section~\ref{ashbt}
with determining the correlation radii. Finally, in Section~\ref{strategy} we present
a simple strategy how various parameters of an azimuthally non-symmetric fireball
can be obtained from data. We conclude in Section~\ref{conc}.

%%%%%%%%%%%%%%%%%%%%%%%%%%%%%%%%%%%%%%%%%%%%%%%%%%%%%%%%%%%
\section{Buda-Lund model: basic features}
\label{model}

Perfect fluid hydrodynamics is based on local conservation of
entropy $\sigma$ and four-momentum tensor $T^{\mu\nu}$,
\begin{eqnarray}
    \partial_\mu (\sigma u^\mu) & = & 0, \\
    \partial_\nu T^{\mu \nu} & = & 0,
\end{eqnarray}
where $u^\mu$ stands for the four-velocity of the matter, which is normalized
to unity as $u^\mu u_\mu = 1$.
(Entropy conservation indicates that there are no dissipative terms
like bulk or shear viscosity, and heat conduction, as any of these
effects would lead to entropy production).
For perfect fluids, the four-momentum tensor is diagonal in the
local rest frame,
\begin{equation}
    T^{\mu \nu}  =  (e + p) u^\mu u^\nu - p g^{\mu \nu}.
\end{equation}
Here $e$ stands for the local energy density and $p$ for
the pressure. The local conservation  of energy, momentum and entropy
yield five constraints on six unknowns (the three independent components
of the four-velocity field and the three thermodynamical variables,
the energy density, the entropy density and the pressure).
These equations are closed by the equation(s) of state,
which gives the relationship between $e$, $p$ (and $\sigma$),
sometimes as a function of yet another variables, like the temperature $T$.

We focus here on the analytic approach in exploring the
consequences of the presence of such perfect fluids in high energy
heavy ion experiments in Au+Au collisions at RHIC. Such exact
analytic solutions were published recently in
refs.~\cite{Csorgo:2001xm,Csorgo:2003rt,Sinyukov:2004am,Csorgo:2006ax,Csorgo:2003ry}.
In some of the recently found exact analytic solutions of
relativistic hydrodynamics, the equation of state has a relatively
simple form, typically $e - B= \kappa (p+B)$ is assumed,
where $B$ is a constant, called the
bag constant which has a non-zero value in a quark-gluon plasma
and has a vanishing value in the hadron gas phase. The parameter $\kappa$
is  either a constant like in ref.~\cite{Csorgo:2003ry} or more
generally, analytic solutions can be obtained even if $\kappa =
\kappa(T)$ is  an arbitrary, temperature dependent function, see
for example the non-relativistic exact solutions in
ref.~\cite{Csorgo:2001xm}.

A tool, that includes these above listed exact, dynamical hydro
solutions, as well as some other cases, and interpolates among them,
 is  the Buda-Lund hydro model of refs.
~\cite{Csorgo:1995bi,Csanad:2003qa}.  This hydro model is
successful in describing experimental data on single particle
spectra and two-particle correlations
~\cite{Csanad:2004cj,Csanad:2003sz}.

The Buda-Lund hydro model~\cite{Csorgo:1995bi} successfully
describes BRAHMS, PHENIX, PHOBOS and STAR data on identified
single particle spectra and the transverse mass dependent
Bose-Einstein or HBT radii as well as the pseudorapidity
distribution of charged particles in central Au + Au collisions
both at $\sqrt{s_{\rm NN}} = 130$ GeV~\cite{Csanad:2003sz} and at
$\sqrt{s_{\rm NN}} = 200$ GeV~\cite{Csanad:2004cj} and in p+p
collisions at $\sqrt{s} = 200$ GeV~\cite{Csorgo:2004id}, as well
as data from Pb+Pb collisions at CERN SPS~\cite{Ster:1999ib} and
$h+p$ reactions at CERN SPS~\cite{Csorgo:1999sj,Agababyan:1997wd}.
The model is defined with the help of its emission function; to
take into account the effects of long-lived resonances, it
utilizes the core-halo model~\cite{Csorgo:1994in}.

In the Buda-Lund model, the emission function is given by that of
a hydrodynamically expanding fireball, surrounded by a halo of
long lived resonances. In this paper, we assume the validity of
the core-halo model, i.e. that the long-lived resonances create an
unresolved, narrow peak in the two-particle HBT correlation
function, which results in an effective intercept parameter
$\lambda$. The integrals of the emission function of the
hydrodynamically expanding core are evaluated~\cite{Csorgo:1999sj}
in a saddle-point approximation. When an improved saddle-point
calculation is performed, effective binary sources and related
non-Gaussian behavior is seen in this model, in particular in the
longitudinal direction~\cite{Csorgo:1999sj}. However, in a
Gaussian approximation, using a single saddle point in the
integration, the effective Buda-Lund emission function looks like:
\begin{eqnarray}\label{e:blsource}
S(x,k) d^4 x & = & \frac{g}{(2 \pi)^3} \frac{p^\mu u_\mu(x_f)\,
H(\tau_f) }{B(x_f,p)+s_q} \times\\
& \times & \exp\left[-R^{-2}_{\mu \nu} (x-x_f)^\mu(x-x_f)^\nu
\right]d^4x,\nonumber
\end{eqnarray}
where $g$ is the degeneracy factor ($g = 1$ for identified
pseudoscalar mesons, $g = 2$ for identified spin=1/2 baryons),
$B(x,p)$ is the (inverse) Boltzmann phase-space distribution, the
term $s_q$ is determined by quantum statistics, $s_q = 0$, $-1$,
and $+1$ for Boltzmann, Bose-Einstein and Fermi-Dirac
distributions, respectively, and
the space-time four-vector of the point of maximal emissivity
(or freeze-out) $x_f \equiv  x_f(p)$ depends on the momentum of the
emitted particles.  The time dependence
of the emission is described by $H(\tau)$, and
\begin{equation}
R^{-2}_{\mu \nu} = \partial_{\mu}\partial_{\nu} (-
\ln(S_{0}))(x_f,p),
\end{equation}
is the 'width' of the emission function, introducing $S_{0}$, as
the 'main' part of the emission function:
\begin{equation}
S_{0}(x,p)= \frac{H(\tau)}{B(x,p)+s_q}.
\end{equation}

For a relativistic, hydrodynamically expanding system, the
(inverse) Boltzmann phase-space distribution is
\begin{equation}
B(x,p)=\exp\left( \frac{ p \cdot u(x)}{T(x)} -\frac{\mu(x)}{T(x)}
\right).
\end{equation}
We will utilize some ansatz for the shape of the flow
four-velocity, $u_\nu(x)$, chemical potential, $\mu(x)$, and
temperature, $T(x)$ distributions. Their form is determined with
the help of exact solutions of hydrodynamics, both in the
relativistic~\cite{Csorgo:2003rt,Csorgo:2003ry,Csorgo:2006ax} and
in the non-relativistic
cases~\cite{Csorgo:2001ru,Csorgo:1998yk,Csorgo:2002kt,Akkelin:2000ex},
with the conditions that these distributions are characterized by
mean values and variances, and that they lead to (simple) analytic
formulas when evaluating particle spectra and two-particle
correlations.

The solution is scale-invariant with the following scaling
variable:
\begin{equation}
s=\frac{r_x^2}{X^2}+\frac{r_y^2}{Y^2}+\frac{r_z^2}{Z^2}.
\end{equation}

The velocity field is
\begin{equation}
u(x)=(\gamma, r_x \frac{\dot X}{X}, r_y \frac{\dot Y}{Y}, r_z
\frac{\dot Z}{Z}),
\end{equation}
with $\gamma$ being the appropriate factor to ensure $u\cdot u=1$.

For the fugacity distribution we assume a shape, that leads to
Gaussian profile in the non-relativistic limit,
\begin{equation}
\frac{\mu(x)}{T(x)} = \frac{\mu_0}{T_0} - s,
\end{equation}
corresponding to the solution discussed in
refs.~\cite{Csorgo:2001ru,Csorgo:1998yk,Csorgo:2001xm}. We assume
that the temperature may depend on the position as well as on
proper-time. We characterize the inverse temperature distribution
similarly to the shape used in the axially symmetric model of
ref.~\cite{Csorgo:1995bi,Csorgo:1995vf}, and discussed in the
exact hydro solutions of refs~\cite{Csorgo:2001ru,Csorgo:1998yk},
\begin{equation}
\frac{1}{T(x)}= \frac{1}{T_0} \left( 1 + \frac{T_0 - T_s}{T_s}
\:s\right) \left( 1 + \frac{T_0 - T_e}{T_e} \, \frac{(\tau
-\tau_0)^2}{2 \Delta\tau^2}\right),
\end{equation}
where $T_0$ and $T_s$ are the temperatures of the center,
and the surface at the mean freeze-out time $\tau_0$, while $T_e$
corresponds to the temperature of the center after most of the
particle emission is over (cooling due to evaporation and
expansion). Sudden emission corresponds to  $T_e = T_0$, and the
$\Delta\tau \rightarrow 0$ limit. It is convenient to introduce the
following quantities
\begin{eqnarray}
\label{atemp}
a^2 & = & \frac{T_0 - T_s}{T_s} = \left< \frac{\Delta T}{T} \right>_\bot, \\
d^2 & = & \frac{T_0 - T_e}{T_e} = \left< \frac{\Delta T}{T} \right>_\tau
\end{eqnarray}

The time dependence of the emission, described by $H(\tau)$ can be
approximated with a Gaussian representation of the Dirac-delta
distribution around the freeze-out proper-time $\tau_f$,
\begin{equation}
H(\tau) = \frac{1}{(2 \pi \Delta\tau^2)^{1/2}}
\exp\left(-\frac{(\tau - \tau_f)^2 }{ 2 \Delta \tau^2}\right),
\end{equation}
with $\Delta \tau$ being the proper-time duration of the particle
production.

From now on our investigations will be performed at {\em mid\-rapidity}.
With the above shorthands, at midrapidity the Boltzmann factor at the freeze-out will have the
following form:
\begin{equation}
B(x_f,p)= \exp \left(\frac{p_x^2}{2 m_t T_x} + \frac{p_y^2}{2 m_t
T_y} - \frac{p_t^2}{2 m_t T_0} + \frac{m_t}{T_0} -
\frac{\mu_0}{T_0}\right),
\end{equation}
where $m_t = \sqrt{p_t^2 + m^2}$
and the direction
dependent slope parameters are
\begin{eqnarray}\label{e:tstar}
T_x&=&T_0+m_t \, \dot X^2 \frac{T_0}{T_0 + m_t a^2},\\
T_y&=&T_0+m_t \, \dot Y^2 \frac{T_0}{T_0 + m_t a^2}.
\end{eqnarray}

%%%%%%%%%%%%%%%%%%%%%%%%%%%%%%%%%%%%%%%%%%%%%%%%%%%%%%%%%%%%

\section{Elliptic flow}
\label{vtwo}
The result for the elliptic flow, that comes directly from a
perfect hydro solution is the following simple scaling
law~\cite{Csorgo:2001xm,Csanad:2003qa}
\begin{equation}
\label{vteq}
v_2=\frac{I_1(w)}{I_0(w)},
\end{equation}
where $I_n(z)$ stands for the modified Bessel function of the
second kind, $I_n(z) = (1/\pi) \int_0^\pi \exp(z \cos(\theta))
\cos(n \theta) d\theta $. The scaling variable
\begin{equation}
w = \frac{p_t^2}{4m_t} \left ( \frac{1}{T_y} - \frac{1}{T_x} \right )\, .\label{e:w}
\end{equation}
This can also be written as
\begin{equation}
w = E_K\, \frac{\epsilon}{T_{\rm eff}}\, ,
\end{equation}
where $E_K$ is a relativistic generalization of the transverse
kinetic energy, defined as
\begin{equation}
E_K = \frac{p_t^2}{2 m_t}\, ,
\end{equation}
and we have introduced $T_{\rm eff}$, the effective slope of the azimuthally
averaged single particle $p_t$ spectra as the harmonic mean of the
slope parameters in the in-plane and in the out-of-plane transverse directions,
\begin{equation}
T_{\rm eff} = 2 \left ( \frac{1}{T_x} + \frac{1}{T_y} \right )^{-1}\, ,\label{e:teff}
\end{equation}
and  momentum space eccentricity parameter
\begin{equation}
\epsilon  =  \frac{T_x - T_y}{T_x + T_y}\, .
\end{equation}
If there is no transverse temperature gradient ($a=0$), then this simplifies as
\begin{equation}
\label{epsdeft}
\epsilon  =  \frac{\dot X^2 - \dot Y^2}{\dot X^2 + \dot Y^2 + 2 T_0/m_t}\, ,
\end{equation}
(see refs.~\cite{Csorgo:2001xm,Csanad:2003qa,Csanad:2005gv} for details).

Thus the Buda-Lund hydro model
predicted~\cite{Csorgo:2001xm,Csanad:2003qa} a \emph{universal
scaling} of the elliptic flow: every $v_2$ measurement
is predicted to fall on the same
scaling curve $I_1/I_0$ when plotted against the scaling variable
$w$. This means, that $v_2$ depends on any physical parameter
(transverse or longitudinal momentum, mass, center of mass energy,
collision centrality, type of the colliding nucleus etc.) only
through $w$.

We note, that for a
leading order approximation, $E_K \approx m_t - m$, which also
explains recent development on scaling properties of $v_2$ by the
PHENIX experiment at
midrapidity~\cite{Adare:2006ti,Afanasiev:2007tv}.

In general, the azimuthal anisotropy of hadronic momentum spectrum is
generated by the azimuthal variation of both the  absolute magnitude of the expansion
velocity and its orientation. The same elliptic flow parameter $v_2$ can be obtained
by different combinations of these two and it has been shown in the
framework of blast-wave model that by modifying the latter one can completely change
(even invert) the dependence of $v_2$ on the former~\cite{Tomasik:2004bn}.
Technically, this shows up as  a dependence of the elliptic flow on both the spatial anisotropy
and the flow anisotropy.

We see from eqs.~\eqref{vteq}--\eqref{epsdeft} that in Buda-Lund model the
velocity field is oriented in such a way that the  spatial anisotropy of the hadronic
freeze-out stage plays no role in determining $v_2$:
eq.~\eqref{vteq} depends only on the transverse expansion rates $\dot X$ and $\dot Y$,
but it does not depend on the actual source sizes $X$ and $Y$.
Only difference of the slope parameters $T_x$ and $T_y$ is
important for the elliptic flow. Thus the elliptic flow vanishes in this model,
if $\dot X = \dot Y$, indicating that if the expansion rates are similar in the
in-plane and out-of-plane transverse directions, $v_2 = 0$. In this sense,
we may write that $v_2$ in the Buda-Lund model is driven by the difference
of the in-plane and out-of-plane transverse expansion rates of the fluid,
during the hadronic final state.

In order to compare more easily the Buda-Lund model results for the elliptic flow with the
azimuthally sensitive extension of the blast-wave model \cite{Retiere:2003kf,Tomasik:2004bn},
we introduce $\rho_0$ and $\rho_2$ so that
\begin{subequations}
\begin{eqnarray}
\dot X & = & \rho_0 (1 + \rho_2)\\
\dot Y & = & \rho_0 (1 - \rho_2)
\end{eqnarray}
\end{subequations}
therefore
\begin{subequations}
\begin{eqnarray}
\rho_0 & = & \frac{1}{2} \left ( \dot X + \dot Y\right )\\
\rho_2 & = & \frac{\dot X - \dot Y}{\dot X + \dot Y}\, .
\end{eqnarray}
\end{subequations}

\begin{figure}[t]
  \begin{center}
  \includegraphics[width=0.62\linewidth,angle=-90]{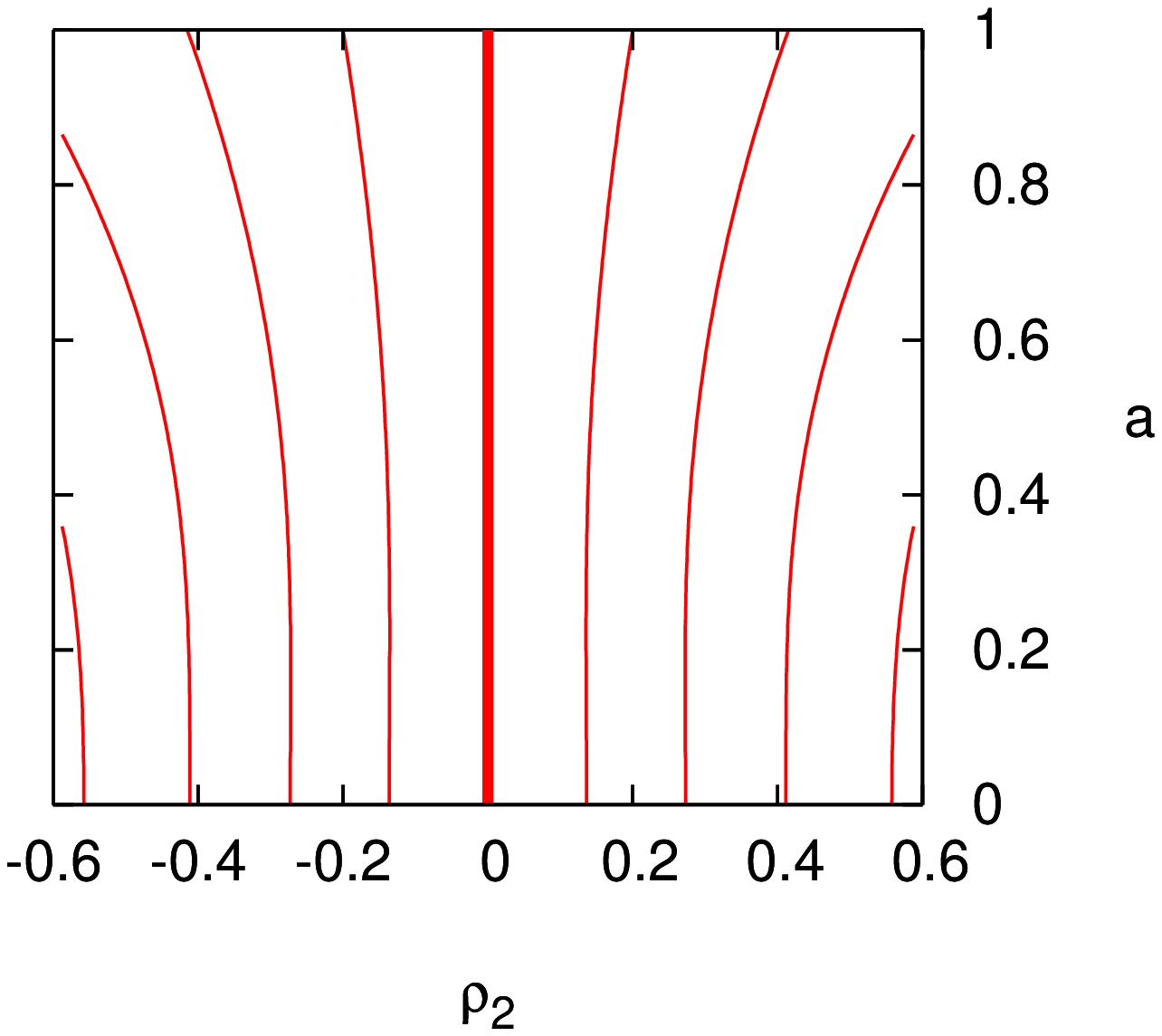}
  \includegraphics[width=0.62\linewidth,angle=-90]{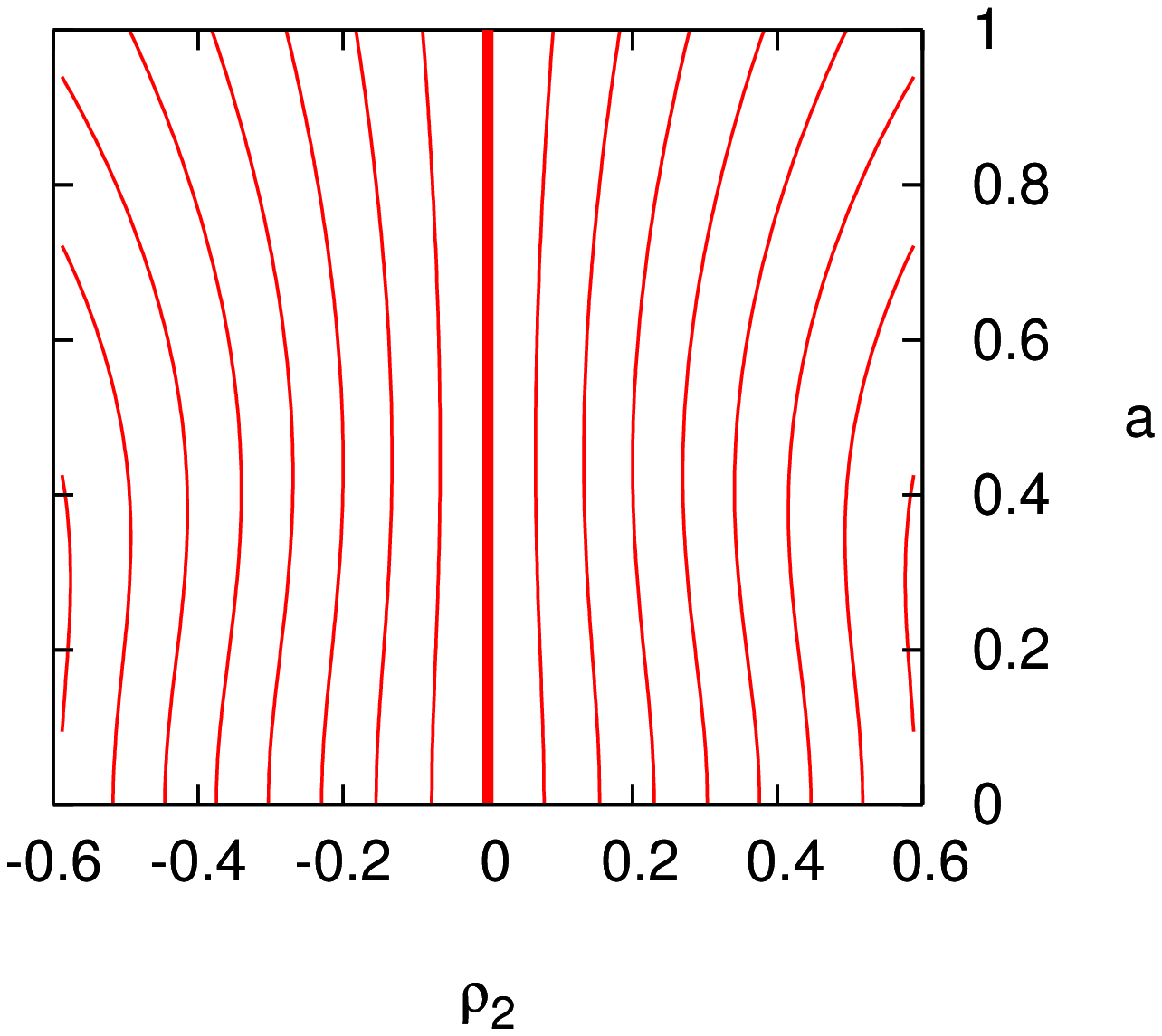}
  \end{center}
  \caption{Lines of constant $v_2$ of pions as function of $\rho_2$ and $a$.
  The increment between
  neighboring contours is 0.05, the thick curve shows $v_2 = 0$. The sign of $v_2$
  is the same as the sign of $\rho_2$.
  Transverse momentum is 0.5~GeV/$c$ (top) and 1~GeV/$c$ (bottom). Other parameters
  in the calculation were
  motivated by data (see Section~\ref{strategy}): $T_0 = 163$~MeV, $\rho_0 = 0.61$.
  \label{f:v2contour}}
\end{figure}

The elliptic flow in the Buda-Lund model also depends on the
transverse temperature gradient $a$ of eq.~(\ref{atemp}). It
actually moderates the difference between $T_x$ and $T_y$ and
increases the elliptic flow at given $\rho_2$. The dependence of
$v_2$ on flow anisotropy $\rho_2$ and temperature gradient $a$ is
shown in Fig.~\ref{f:v2contour}.
%%%%%%%%%%%%%%%%%%%%%%%%%%%%%%%%%%%%%%%%%%%%%%%
%%%%%%%%%%%%%%%%%%%%%%%%%%%%%%%%%%%%%%%%%%%%%%%%%%%%%
\section{Azimuthally dependent correlation radii}
\label{ashbt}

The two-particle Bose-Einstein correlation
function from the Buda-Lund source function (defined in eq.~\eqref{e:blsource})
is most naturally calculated in the frame given by the main axes of the ellipsoid.
The result is
\begin{equation}
C(q)=1 + \lambda {\rm e}^{-q_0^2 \Delta \tau_{*}^2 - q_x^2
R_{*,x}^2 - q_y^2 R_{*,y}^2 - q_z^2 R_{*,z}^2}\label{e:corr}
\end{equation}
with $\lambda$ being the intercept parameter, $q = p_1 -  p_2$, and
\begin{subequations}
\label{starrad}
\begin{eqnarray}
\frac{1}{\Delta\tau_*^2} &=& \frac{1}{\Delta \tau^2} + \frac{m_t}{T_0} \frac{d^2}{\tau_0^2},\\
R_{*,x}^2&=&X^2\left(1+{\frac{m_t\, \left(a^2+\dot X^2 \right)}{T_0}}\right)^{-1}, \label{e:Rx}\\
R_{*,y}^2&=&Y^2\left(1+{\frac{m_t\, \left(a^2+\dot Y^2 \right)}{T_0}}\right)^{-1}, \label{e:Ry}\\
R_{*,z}^2&=&Z^2\left(1+{\frac{m_t\, \left(a^2+\dot Z^2 \right)}{T_0}}\right)^{-1}.
\end{eqnarray}
\end{subequations}
Note that parameter $d$ makes the effective life-time and the difference between
the out and the side radius components momentum dependent.

From the mass-shell constraint of $p_1^2 = p_2^2 = m^2$ one finds that
\begin{equation}
q_0 = \beta_x q_x + \beta_y q_y + \beta_z q_z
\end{equation}
where $\vec \beta = \vec K/K^0 = (\vec p_1 - \vec p_2)/(p_1^0 - p_2^0)$.
Thus we can rewrite eq.~\eqref{e:corr} to
\begin{equation}
C(q)=1 + \lambda \exp\left(-\sum_{i,j=x,y,z} R^2_{ij} q_i
q_j\right)
\end{equation}
with the modified radii of
\begin{subequations}
\begin{eqnarray}
R_x^2 & = & R_{*,x}^2 + \beta_x^2 \Delta \tau_{*}^2,\\
R_y^2 & = & R_{*,y}^2 + \beta_y^2 \Delta \tau_{*}^2,\\
R_z^2 & = & R_{*,z}^2 + \beta_z^2 \Delta \tau_{*}^2,\\
R_{xz}^2 & = & \beta_x \beta_z \Delta \tau_{*}^2,\\
R_{yz}^2 & = & \beta_y \beta_z \Delta \tau_{*}^2,\\
R_{xy}^2 & = & \beta_x \beta_y \Delta \tau_{*}^2.
\end{eqnarray}
\end{subequations}
The above formulas are given in the frame, specified by the main axes of the expanding,
 ellipsoidally symmetric fireball.
%This system
%is called the System of Ellipsoidal Expansion (SEE) introduced
%in ref.~\cite{Csorgo:2001xm}.
In general, this frame is rotated~\cite{Csorgo:2001xm}  with respect to
the standard out-side-long system~\cite{Pratt:1986cc},
which is given by the beam direction (longitudinal) and the
transverse component of the pair momentum (outward). The
correlation function in this out-side-long system is parameterized as
\begin{equation}
C(q) = 1+ \lambda {\rm e}^{ - q_o^2 R_o^2 - q_s^2 R_s^2 - q_l^2
R_l^2 - 2(q_oq_sR_{os}^2+q_oq_lR_{ol}^2+q_sq_l R_{sl}^2) }\, .
\end{equation}
We denote by
$\vartheta$ the tilt angle between the longitudinal direction and the $z$-axis and $\varphi$ stands for the angle between the outward direction and the plane given by $x$ and $z$ axes---the
reaction plane. In the new frame we obtain azimuthally sensitive HBT  radii that are given as
\begin{subequations}
\begin{eqnarray}
R_o^2 & = & R_x'^2 \cos ^2 \varphi+R_y^2 \sin^2 \varphi +  R_{xy}'^2 \sin (2\varphi),\\
R_s^2 & = & R_x'^2 \sin ^2 \varphi+R_y^2 \cos^2 \varphi -  R_{xy}'^2 \sin(2\varphi),\\
R_l^2 & = & R_x^2 \sin^2 \vartheta + R_z^2 \cos^2 \vartheta +  R_{xz}^2 \sin(2\vartheta),\\
2 R_{os}^2 & = & -R_x^2 \sin(2\varphi)+ R_y^2 \sin(2\varphi)+ 2 R_{xy}'^2 \cos (2\varphi),\\
2 R_{sl}^2 & = & \left(R_x^2\sin(2\vartheta) + R_z^2\sin(2\vartheta)-2R_{xz}\cos(2\vartheta)\right) \sin\varphi \nonumber\\
& & +  \left(2R_{xy}^2\sin \vartheta+2R_{yz}^2 \cos \vartheta\right)\cos\varphi, \\
2 R_{lo}^2 & = & \left(R_x^2\sin(2\vartheta) - R_z^2\sin(2\vartheta) +2R_{xz}\cos(2\vartheta)\right) \cos\varphi \nonumber\\
& & +  \left(2R_{xy}^2\sin \vartheta+2R_{yz}^2 \cos \vartheta\right)\sin\varphi,
\end{eqnarray}
\end{subequations}
where we have introduced
\begin{eqnarray}
R_x'^2 &=& R_x^2 \cos^2 \vartheta + R_z^2 \sin^2 \vartheta -  R_{xz}^2 \sin(2 \vartheta) \\
R_{xy}'^2&=&R_{xy}^2 \cos \vartheta - R_{yz}^2 \sin \vartheta
\end{eqnarray}
If the fireball is not tilted in the reaction plane ($\vartheta = 0$), which might appear at very
high collision energies, then in the
longitudinally co-moving frame (LCMS, $\beta_l = 0$) the formulas further simplify as
\begin{subequations}
\label{oslradii}
\begin{eqnarray}
R_o^2 & = & R_{*,x}^2 \cos ^2 \varphi + R_{*,y}^2 \sin^2 \varphi + \beta_o^2 \Delta \tau_{*}^2 \\
& = &\frac{R_{*,x}^2 + R_{*,y}^2}{2} + \beta_o^2 \Delta \tau_{*}^2
- \frac{R_{*,y}^2 - R_{*,x}^2}{2} \cos(2\varphi)\, \nonumber \\
R_s^2 & = & R_{*,x}^2 \sin ^2 \varphi + R_{*,y}^2 \cos^2 \varphi\\
\label{rside}
& = &\frac{R_{*,x}^2 + R_{*,y}^2}{2} + \frac{R_{*,y}^2 - R_{*,x}^2}{2} \cos(2\varphi)\, \nonumber,\\
R_{os}^2 & = & \frac{ R_{*,y}^2 - R_{*,x}^2 }{2} \sin (2\varphi)\, , \\
R_l^2 & = & R_{*,z}^2\, ,\\
R_{ol}^2 & = & 0 \, , \\
R_{sl}^2 & = & 0\, .
\end{eqnarray}
\end{subequations}
It is convenient to decompose the correlation radii into Fourier series in azimuthal angle
\cite{Heinz:2002au,Tomasik:2002rx}.
Here, it is particularly trivial since there are only zeroth and second order terms
\begin{subequations}
\begin{eqnarray}
R_o^2 & = & R_{o,0}^2 + R_{o,2}^2 \cos(2\varphi)\\
R_s^2 & = & R_{s,0}^2 + R_{s,2}^2 \cos(2\varphi)\\
R_l^2 & = & R_{l,0}^2\\
R_{os}^2 & = & R_{os,0}^2 + R_{os,2}^2 \sin(2\varphi)\, .
\end{eqnarray}
\end{subequations}

In this Gaussian single saddle point version of
the Buda-Lund model there is no implicit azimuthal dependence of the homogeneity regions. Thus the complete dependence on $\varphi$ is
written in eqs.~\eqref{oslradii}. In such a case we observe rather
known features:
\begin{description}
\item{--} The oscillation amplitudes $|R_{o,2}^2|$, $|R_{s,2}^2|$, and
$|R_{os,2}^2|$ are equal.
\item{--} The effective duration of particle emission can be
obtained from the difference of non-oscillating parts  as
\begin{equation}
\beta_o^2 \Delta \tau_{*}^2 = R_{o,0}^2 - R_{s,0}^2\, \label{e:effdur};
\end{equation}
\item{--} We can extract the effective source sizes $R_{*,x}^2$ and $R_{*,y}^2$ from data as
\begin{subequations}
\label{rxry}
\begin{eqnarray}
R_{*,x}^2 & = & R_{s,0}^2 - R_{s,2}^2 ,\\
R_{*,y}^2 & = & R_{s,0}^2 + R_{s,2}^2\, .
\end{eqnarray}
\end{subequations}
\end{description}
If the first feature is not observed, or, if data lead to a negative
r.h.s. of eq.~(\ref{e:effdur}), this would indicate that the
above presented Gaussian version of the Buda-Lund model is not applicable
for the description of that given data.
It is however well known, already in the first formulation of the Buda-Lund model,
 ref.~\cite{Csorgo:1995bi},
the r.h.s. of eq.~(\ref{e:effdur}) can be negative.
The exact hydro  solutions and symmetry arguments require
that $R_{s,0}=R_{o,0}$
in the vanishing transverse momentum limit only,
otherwise at finite transverse momentum,
$R_{s,0} > R_{o,0}$ is allowed in general. Such a behavior happens
in exact, ring of fire type of hydrodynamical solutions,
which were found in fireball hydrodynamics in ref.~\cite{Csorgo:1998yk}.
Such rings of fire type of exact hydro solutions
appear if the radial flow is moderate and the temperature distribution is
very inhomogeneous, in our present notation if $ a^2 \gg m \rho_0^2/(2 T_0)$,
while nearly Gaussian fireballs appear, if the temperature is nearly homogeneous
and the radial flow is strong, $ a^2 \ll m \rho_0^2/(2 T_0)$.
The data discussed in the present paper correspond to this latter limit.

So if there is a significant negative value for $R_{o,0}^2 - R_{s,0}^2$
in some data set, the shell of fire type of Buda-Lund solutions can be utilized,
as a generalization of the present approach.
as e.g. discussed also in refs.~\cite{Csorgo:1999sj,Csorgo:2001ru}.
Such shell of fire structures are also seen in blastwave models,
see e.g. Figs. 18 and 19 of ref.~\cite{Retiere:2003kf}
and also in cascade calculations e.g. in Fig. 12 of
ref.\cite{Humanic:2005ye}.

The azimuthal dependence of correlation radii may be due to
spatial deformation of the fireball as well as due to azimuthal
dependence of the expansion velocity field. The interplay of these
two effects in framework of the blast-wave model has been studied
in \cite{Tomasik:2004bn}. In order to compare with that paper we
parameterize the transverse scales with the help of average radius
$R$ and asymmetry parameter $a_s$
\begin{equation}
\label{XYas}
X = a_s\, R \qquad \qquad Y = \frac{R}{a_s}\, ,
\end{equation}
(note that $a_s$ corresponds to $a$ of \cite{Tomasik:2004bn}).

The amplitude of oscillation trivially scales with the absolute
size of the fireball. We want to scale out this effect in order to
directly see the effect of eccentricity. Thus we shall be
interested in {\em scaled} oscillation amplitudes
$R_{s,2}^2/R_{s,0}^2$. (For the outward radius the result would be
similar---scaled amplitude would be just slightly smaller due to
added $\beta_o^2\Delta\tau_{*}^2$ term in the denominator.) Such a
ratio then does not depend on $R$ and from eqs.\eqref{starrad},
\eqref{rside}, and \eqref{XYas} we derive
\begin{equation}
\label{soa}
\frac{R_{s,2}^2}{R_{s,0}^2} = \frac{\frac{{a_s^{-2}}}{1 + {m_t \, T_0^{-1}\, (a^2 + \rho_0^2(1-\rho_2)^2)}  }
- \frac{a_s^2}{1 + {m_t\, T_0^{-1}\,  (a^2 + \rho_0^2(1+\rho_2)^2)}  }}%
{\frac{{a_s^{-2}}}{1 + {m_t \, T_0^{-1}\, (a^2 + \rho_0^2(1-\rho_2)^2)}  }
+ \frac{a_s^2}{1 + {m_t\, T_0^{-1}\,  (a^2 + \rho_0^2(1+\rho_2)^2)}  }}
\end{equation}
%%%%%%%%%%%%%%%%%%%%%%%%%%%%%%%%%%%%%%%%%%%%%%%%
\begin{figure}[t]
  \begin{center}
  \includegraphics[width=0.62\linewidth]{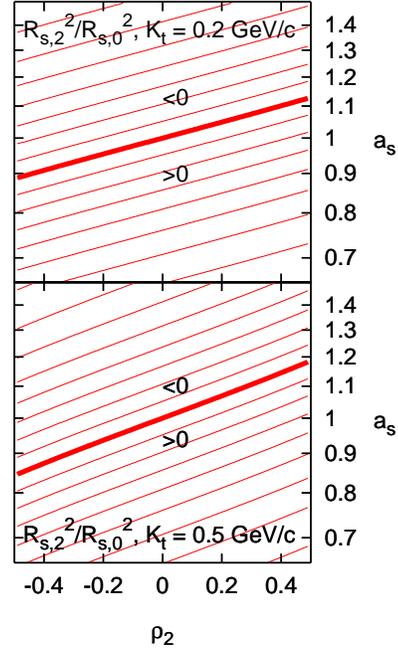}
  \end{center}
  \caption{%
  Curves of constant $R_{s,2}^2/R_{s,0}^2$ calculated as functions of
  $a_s$ and $\rho_2$ at fixed $a=0$. Temperature $T_0 = 163$~MeV and
  $\rho_0 = 0.61$ (as obtained in Section~\ref{strategy}). Increments between
  the contours are 0.1, thick lines show $R_{s,2}^2/R_{s,0}^2=0$. Pair momentum
  $K_t$ was chosen 0.2~GeV (upper panel) and 0.5~GeV (lower panel).
  \label{f:rsat0}}
\end{figure}
%%%%%%%%%%%%%%%%%%%%%%%%%%%%%%%%%%%%%%%%%%%%%%%
\begin{figure}[t]
  \begin{center}
  \includegraphics[width=0.62\linewidth]{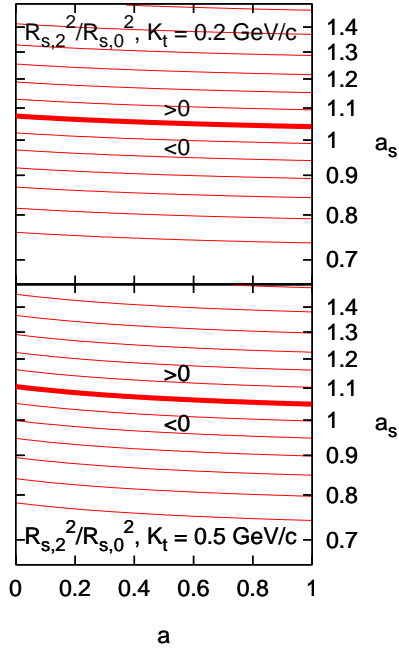}
  \end{center}
  \caption{%
  Curves of constant $R_{s,2}^2/R_{s,0}^2$ calculated as functions of
  $a_s$ and $a$ at fixed $\rho_0=0.3$. Temperature $T_0 = 163$~MeV and
  $\rho_0 = 0.61$ (as obtained in Section~\ref{strategy}). Increments between
  the contours are 0.1, thick lines show $R_{s,2}^2/R_{s,0}^2=0$. Pair momentum
  $K_t$ was chosen 0.2~GeV (upper panel) and 0.5~GeV (lower panel).
  \label{f:rsrh3}}
\end{figure}
%%%%%%%%%%%%%%%%%%%%%%%%%%%%%%%%%%%%%%%%%%%%%%%%%%%%%
We plot these scaled oscillation amplitudes in Figure~\ref{f:rsat0} as functions
of $a_s$ and $\rho_2$. The correlation between spatial and flow asymmetry is well pronounced:
one can get the same oscillation with different pairs of values for the
two parameters. Note that the dependence on flow anisotropy $\rho_2$ in the
Buda-Lund model is much stronger then in the blast-wave model of
refs.~\cite{Retiere:2003kf} and \cite{Tomasik:2004bn} (Model 1 of the latter paper).

There is an additional parameter in the Buda-Lund model, which
influences the oscillation: the transverse temperature gradient
$a$ defined in eq.~(\ref{atemp}). We plot the dependence of scaled
oscillation amplitude on $a$ and $a_s$ in Figure~\ref{f:rsrh3}.
From eq.~\eqref{soa} one can infer that the amplitude does not
depend of $a$ if $\rho_2=0$. For the plots we have chosen $\rho_2
= 0.3$. Increasing $a$ tends to wash away the dependence of the
amplitude on $\rho_2$, as seen from eq.~\eqref{soa}.
%%%%%%%%%%%%%%%%%%%%%%%%%%%%%%%%%%%%%%%%%%%%%%%%%%%%%%%%%%%%%%%%%%

\section{Strategy of analysis}
\label{strategy}

In this section we outline how one can infer the model parameters from
data on (i) azimuthally integrated spectra, (ii) elliptic flow, (iii)
azimuthally and transverse momentum dependent correlation radii.

For a simple analysis strategy, one might observe,
that approximate formulas can be derived assuming that the
flow anisotropy parameter is small, $\rho_2 \ll 1$.
Keeping the leading order terms in this expansion,
the following approximative formulas are obtained:
\begin{eqnarray}
T_{\rm eff} & = & T_0 + M \rho_0^2+ {\cal{O}}(\rho_2^2) ,\label{e:teffs}\\
M &=&	m_t \frac{T_0 }{T_0 + m_t a^2}, \\
v_2 & \simeq & \frac{p_t^2}{2 m_t M} \frac{\rho_2}{\rho_0^2}\frac{1}{(1 + \frac{T_0}{M \rho_0^2})^2 }.
\end{eqnarray}
Here, $T_{\rm eff}$ is the effective slope parameter of {\em
azimuthally integrated} $m_t - m$ spectrum .
Let us observe,  that this effective temperature depends on the freeze-out
temperature scale $T_0$ and on the radial flow coefficient $\rho_0$,
but to a leading order, it does not depend on the flow anisotropy parameter $\rho_2$.
In contrast, the elliptic flow is increasing linearly with $\rho_2$.
Thus, naively, one expects that the mass dependence of the slope parameters
can be utilized to determine the freeze-out temperature $T_0$ and the radial flow
parameter $\rho_0$, then using these parameters one could extract the
flow anisotropy parameter $\rho_2$ from the data on the elliptic flow $v_2$.

{\it If} the temperature inhomogeneity parameter is
very small, $a^2 \ll T_0/m_t$, then  $M \approx m_t$, and one can adopt such a simple
minded strategy. Let us then consider the HBT radii in this $a \simeq 0$ case:
\begin{eqnarray}
R_s & \simeq & \frac{R_{*,x}^2+R_{*,y}^2}{2} + \frac{R_{*,y}^2-R_{*,x}^2}{2} \cos 2 \varphi \\
R_{*,x} & \simeq & X^2 \frac{T_0}{T_0 + m_t \dot X^2 },\\
R_{*,y} & \simeq & Y^2 \frac{T_0}{T_0 + m_t \dot Y^2}.\label{e:rys}
\end{eqnarray}
As $\dot X$ and $\dot Y$ are given by $\rho_0$ and $\rho_2$ the new fit parameters
will be the actual sizes $X$ and $Y$.
So if $a=0$, one can obtain the freeze-out temperature and the radial flow parameter
from the mass dependence of the slopes, then momentum anisotropy parameter $\rho_2$
can be obtained from  $v_2 $ using $T_0$ and $\rho_0$, finally $X$ and $Y$
come from the HBT radii.
However, the temperature inhomogeneity parameter $a^2$ couples to all expressions
through the effective mass scale $M$ defined above, and makes such a simple strategy
impractical, except in the case of $a= 0$.

As such a naive strategy cannot be applied in the general case,
because parameter $a$ appears in a non-trivial manner in both the slope,
the elliptic flow and the HBT data, we have tried to do a simultaneous description of
slope parameters, elliptic flow and HBT radii.
To estimate a reasonable range of the model parameters, we first have
selected a centrality class, 20-30\%, where
such data are available.

\begin{figure}
  \begin{center}
  \includegraphics[angle=-90,width=1.0\linewidth]{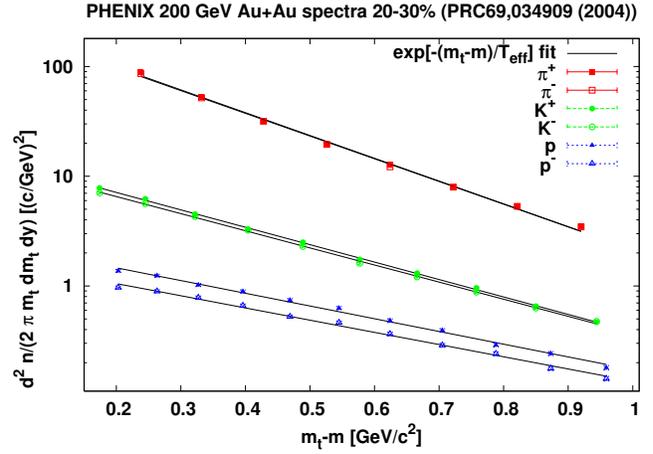}
\caption{The exponential fits to PHENIX
identified pion, kaon, proton (and antiparticle) spectra from
20-30\% centrality class.
%are
%obtained in just the same way as it was done for other centrality classes in
%ref.~\cite{Adler:2003cb}:
%$\frac{d^2n}{(2 \pi m_t dm_t dy)} $ was fitted to
%$\frac{A}{2 \pi T_{\rm eff} (T_{\rm eff}+ m)}\exp(-\frac{m_t-m}{T_{\rm eff}})$. The fit range was
%0.2 GeV $< m_t - m < $ 1 GeV for $\pi^+$ and $\pi^-$,
%while it was 0.1 GeV $< m_t - m < $ for $K^+$, $K^-$, $p$ and $\overline{p}$.
%The effective temperatures are summarized in Table \ref{table:1}.
\label{f-slopes}
}
\end{center}
\end{figure}

First, we have obtained the slopes of single-particle azimuthally integrated spectra
in $m_t - m$. The exponential fits to PHENIX
identified pion, kaon, proton (and antiparticle) spectra from 20-30\% centrality class
are obtained in just the same way as it was done for other centrality classes in
ref.~\cite{Adler:2003cb}:
$\frac{d^2n}{(2 \pi m_t dm_t dy)} $ was fitted to
$\frac{A}{2 \pi T_{\rm eff} (T_{\rm eff}+ m)}\exp(-\frac{m_t-m}{T_{\rm eff}})$.
The fit range was 0.2 GeV $< m_t - m < $ 1 GeV for $\pi^+$ and $\pi^-$,
while it was 0.1 GeV $< m_t - m < $ for $K^+$, $K^-$, $p$ and $\overline{p}$.
The fits are shown in Figure~\ref{f-slopes} and effective temperatures are summarized
in Table \ref{table:1}. PHENIX spectra were used here because they are corrected
for weak decays.
%%%%%%%%%%%%%%%%%%%%%%%%%%%%%%%%%%%%%%%%%%%%%%%%%%%%%%%%%%%%%%%%%%%%
%%%%%%%%%%%%%%%%%%%%%%%%%%%%%%%%%%%%%%%%%%%%%%%%%%%%%%%%%%%%%%%%%
\begin{table}
\begin{center}
\begin{tabular}{|c|c|c|}
  \hline
  \hline
 Particle: & $m$ (MeV) & $T_{\rm eff}$ (MeV)   \\
  \hline
  \hline
 $\pi^+$  	& 139        & 210.2 $\pm$ 0.8  \\ \hline
 $K^+$  	& 494        & 284.4 $\pm$ 2.2  \\ \hline
 $p$  	        & 938        & 414.8 $\pm$ 7.5  \\ \hline
 $\pi^-$  	& 139        & 211.9 $\pm$ 0.7  \\ \hline
 $K^-$  	& 494        & 278.1 $\pm$ 2.2  \\ \hline
 $\overline{p}$ & 938        & 434.7 $\pm$ 8.6  \\ \hline
  \hline
\end{tabular}
\caption{Exponential slope parameters extracted from PHENIX weak decay
	corrected identified particle spectra in 20-30 \% most central  Au+Au  collisions
	at $\sqrt{s_{NN}}= 200$ GeV.  Errors are statistical only.
}\label{table:1}
\end{center}
\end{table}
%%%%%%%%%%%%%%%%%%%%%%%%%%%%%%%%%%%%%%%%%%%%%%%%%%%%%%%%%%%%%%%%%%%%%%%

%%%%%%%%%%%%%%%%%%%%%%%%%%%%%%%%%%%%%%%%%%%%%%%%%%%%%%%%%%%%%%%%%%%%%%
\begin{figure}
  \begin{center}
  \includegraphics[angle=-90,width=0.62\linewidth]{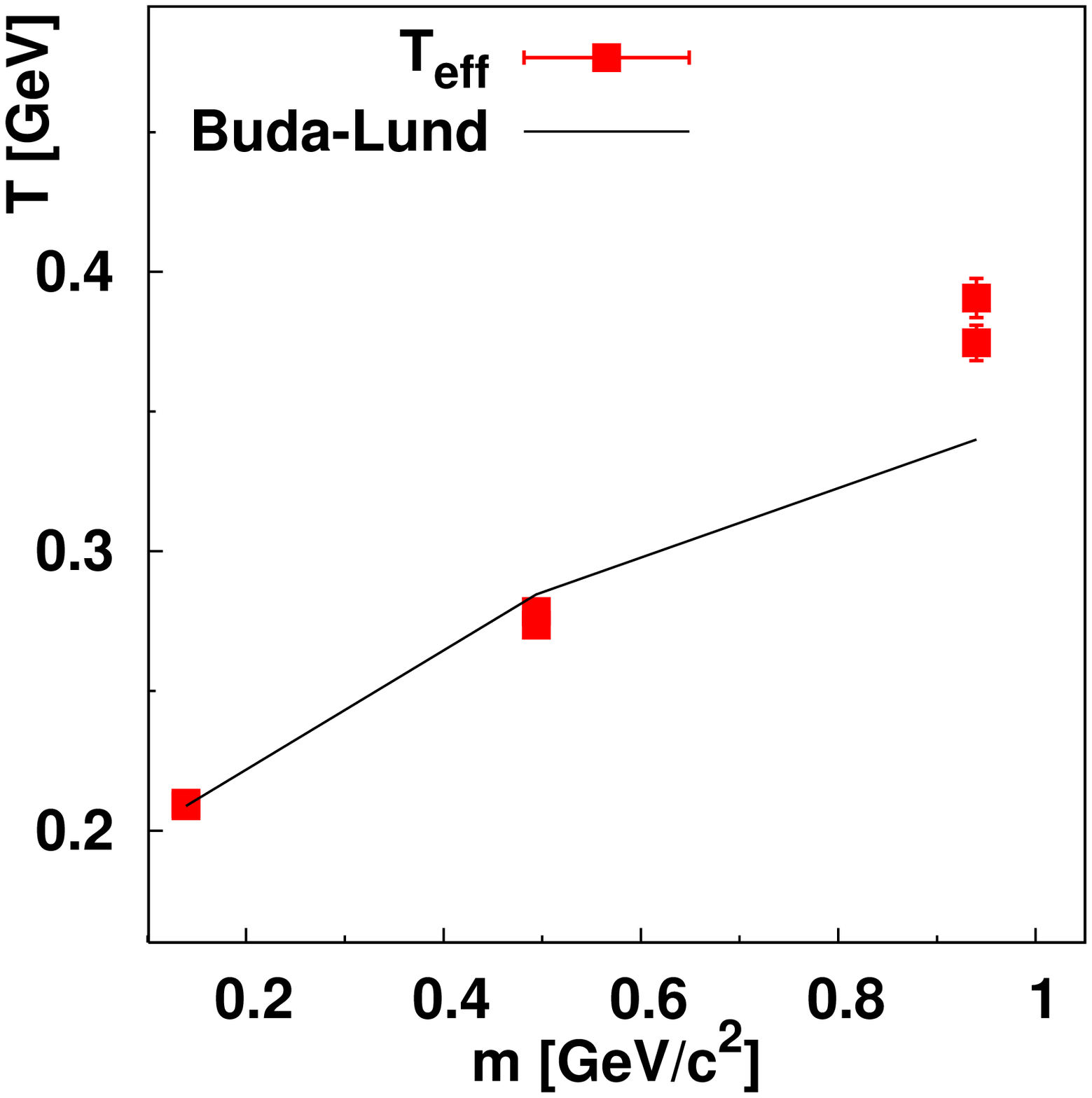}
  \includegraphics[angle=-90,width=0.62\linewidth]{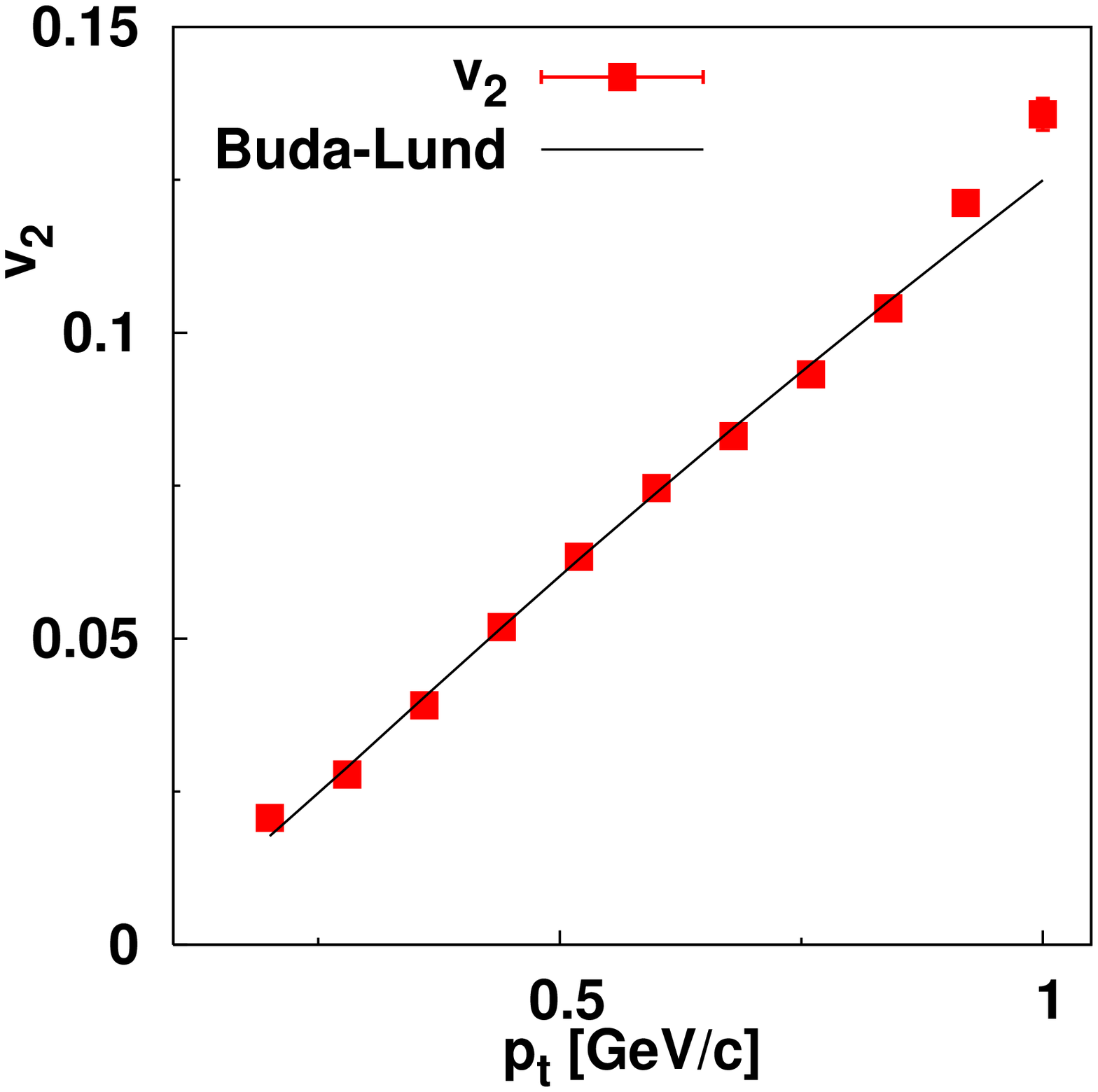}
  \includegraphics[angle=-90,width=0.62\linewidth]{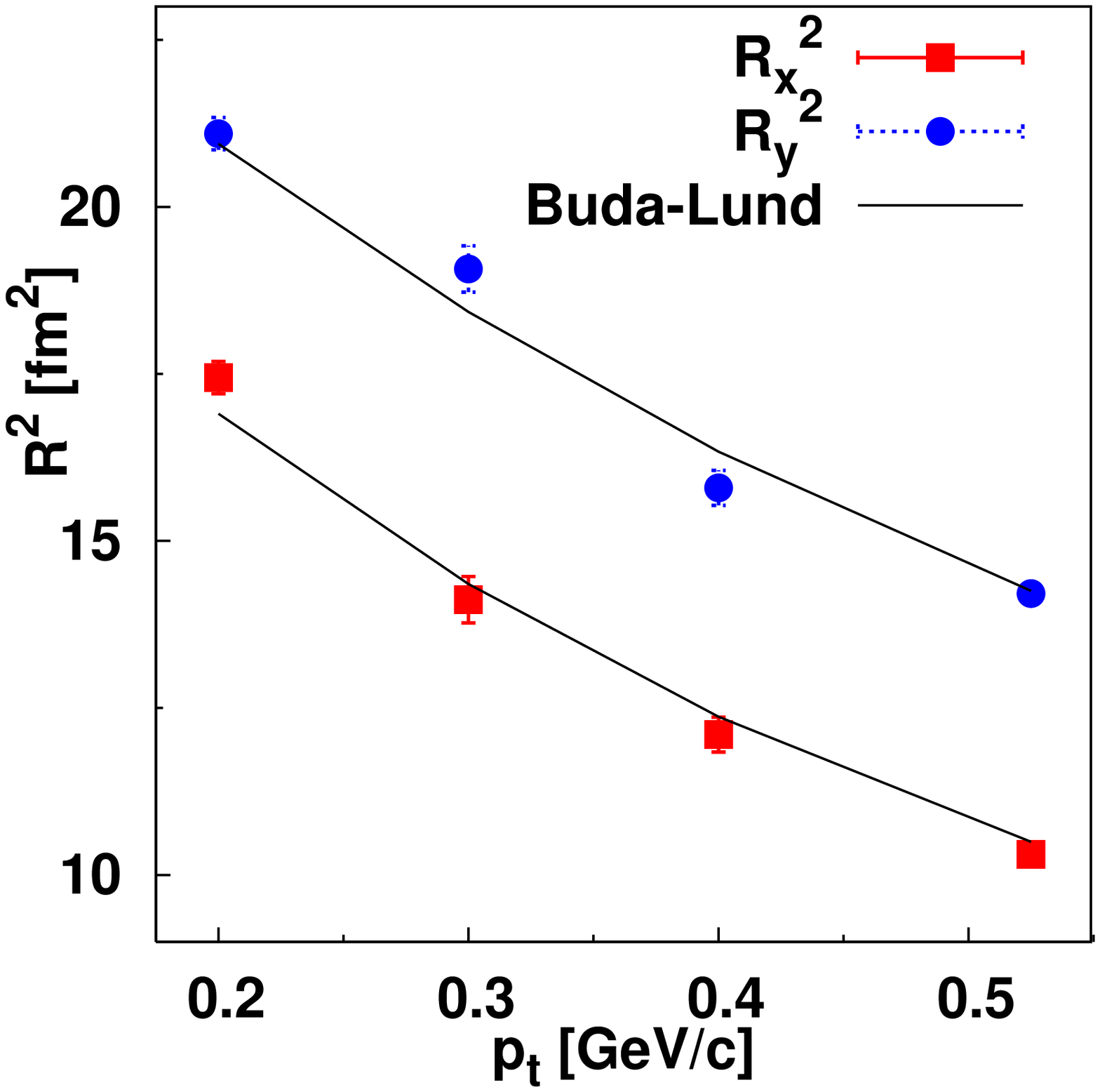}
\caption{
 The top panel shows the
  Buda-Lund calculations compared to  slope parameters obtained from PHENIX
weak decay corrected invariant momentum distribution data~\cite{Adler:2003cb}
in the 20-30\%  centrality class.
The middle panel shows a comparison to  pion $v_2$ data from STAR~\cite{Adams:2004bi}
in the same centrality class.
The bottom panel shows the  Buda-Lund comparison to azimuthally sensitive HBT radii
from STAR~\cite{Adams:2003ra},
also measured in 20-30\% central $\sqrt{s_{\rm NN}} = 200 $GeV Au+Au collisions.
We used the full Buda-Lund formulas for the slopes,
elliptic flow and azimuthally sensitive HBT radii,
eqs.~\eqref{vteq},~\eqref{e:w},~\eqref{e:teff},~\eqref{e:Rx}, and ~\eqref{e:Ry},
and the same parameter values of $T_0 = 163$ MeV, $\rho_0 = 0.61$,
$\rho_2 = 0.17 $, $X^2 = 33.5 $ fm$^2 $, $Y^2 = 33.9$ fm$^2$, and $a^2 = 0.16$
were used in all the three panels.
\label{f:fits} }
  \end{center}
\end{figure}
%%%%%%%%%%%%%%%%%%%%%%%%%%%%%%%%%%%%%%%%%%%%%%%%%%%%%%%%%%%%%%%%%%%%
%With these formulas

In the second step we describe qualitatively the
RHIC 20-30\% central $\sqrt{s_{\rm NN}} = 200 $GeV Au+Au data.
We compared the full Buda-Lund formulas for the slopes,
elliptic flow and azimuthally sensitive HBT radii,
eqs.~\eqref{vteq}, \eqref{e:w}, \eqref{e:teff}, \eqref{e:Rx}, and \eqref{e:Ry} to azimuthally sensitive HBT radii from STAR~\cite{Adams:2003ra},
the mass dependence of effective temperatures from
PHENIX~\cite{Adler:2003cb}, and pion $v_2$ data from
STAR~\cite{Adams:2004bi}. For such an overall procedure, the
 fit parameters are $T_0$, $X$,
$Y$, $\rho_0$, $\rho_2$ and $a$. One can get $R_{*,x}$ and $R_{*,y}$
via eqs.~\eqref{rxry}.
The Buda-Lund calculations are compared
with data in Fig.~\ref{f:fits} for the following parameter
values: $T_0=163$~MeV, $\rho_0=0.61$, $\rho_2= 0.17$, $X^2=33.5$
fm$^2$, $Y^2=33.9$ fm$^2$ and $a^2 = 0.16$.
These results indicate that the theoretical
formulas developed in the present paper are reasonable and
could be utilized to describe qualitatively the azimuthally sensitive data at
in $\sqrt{s_{NN}} = 200 $  GeV Au+Au collisions at RHIC.

%%%%%%%%%%%%%%%%%%%%%%%%%%%%%%%%%%%%%%%%%%%%%%%%%%%%%%%%

\section{Conclusions}
\label{conc}

We have shown that in the Buda-Lund model the expansion of the
velocity field
is oriented in such a way that the elliptic flow only depends on
the asymmetry of the expansion velocity in the directions in and
out of the reaction plane. Unlike the blast-wave model, there is
no influence of the elliptic shape of transverse cross-section.

Oscillation of the transverse correlation radii is shaped by flow
as well as spatial anisotropy. This is again different from the
blast-wave model: in the azimuthally non-symmetric generalization
of that model which was able to describe RHIC data, correlation
radii mainly depend on the ellipsoidal shape and less so on the
anisotropy of the flow.

A first analysis of the single particle spectra, elliptic flow
and azimuthally sensitive HBT radii was presented.
The formulas obtained from the full Buda-Lund model
yield a reasonable description of  these data with a reasonable
choice of the model parameters, indicating, that the ellipsoidally
symmetric Buda-Lund model has reasonable properties when compared
to azimuthally sensitive HBT data.

Note that the present formulas were obtained in a Gaussian
approximation to the proper-time distribution and to the
saddle-point integration of the source function.
Non-Gaussian features were reported earlier when performing
an advanced saddle point integration - deviations from Gaussian
HBT correlations were predicted in the longitudinal direction
already in the azimuthally symmetric version, see
e.g.~\cite{Csorgo:1999sj}.
Also, a non-Gaussian proper-time distribution would result
in non-Gaussian distributions in directions coupled to the
out direction, but these effects are not explored in the current
manuscript.

A unique feature, which could eventually falsify the use of
the single saddle point approximation and the Gaussian version of the Buda-Lund
model for the description of data is the fact that there is only explicit
and no implicit azimuthal dependence of the correlation radii. In such models
oscillation amplitudes of $R_s^2$, $R_o^2$, and $R_{os}^2$ are the same. Precise
data could resolve if this is really the case.

However, some of the conclusions like the fireball freezes out
early, when $X \simeq Y$ but $\dot X > \dot Y$, seem to be robust
features of the data, as these features are obtained in different
models like the Buda-Lund and the blast-wave models independently.

More work has to be done to explore more quantitatively the fitting
procedure, the error estimation and the confidence level tests,
using the full formulas.

%%%%%%%%%%%%%%%%%%%%%%%%%%%%%%%%%%%%%%%%%%%%%%%%%%%%%%%%%%%%%%%%%

\paragraph*{Acknowledgments}
This paper was prepared within a bilateral collaboration
between Hungary and Slovakia under project Nos.\ SK-20/2006 (HU), SK-MAD-02906 (SK).
B.T.\ acknowledges support by
VEGA 1/4012/07 (Slovakia), MSM 6840770039, and
LC 07048 (Czech Republic). M.Cs.\ and T.Cs.\ gratefully acknowledge the support of
the Hungarian OTKA  grants T49466 and NK 73143.
%%%%%%%%%%%%%%%%%%%%%%%%%%%%%%%%%%%%%%%%%%%%%%%%%%%%%%%%%%%%%%%%%%

\bibliographystyle{prlsty}
\bibliography{Master}

\begin{thebibliography}{10}

\bibitem{Riordan:2006df}
M. Riordan and W.~A. Zajc, Sci. Am. {\bf 294N5},  24  (2006).

\bibitem{Retiere:2003kf}
F. Retiere and M.~A. Lisa, Phys. Rev. {\bf C70},  044907  (2004)
  [arXiv:nucl-th/0312024].

\bibitem{Schnedermann:1993ws}
E. Schnedermann, J. Sollfrank, and U.~W. Heinz, Phys. Rev. {\bf C48},  2462
  (1993) [arXiv:nucl-th/9307020].

\bibitem{Csanad:2004cj}
M. Csan\'ad, T. Cs\"org\H{o}, B. L\"orstad, and A. Ster, Nukleonika {\bf 49},
  S49  (2004) [arXiv:nucl-th/0402037].

\bibitem{Csorgo:2001xm}
T. Cs\"org\H{o} {\it et~al.}, Phys. Rev. {\bf C67},  034904  (2003)
  [arXiv:hep-ph/0108067].

\bibitem{Csorgo:2003rt}
T. Cs\"org\H{o}, F. Grassi, Y. Hama, and T. Kodama, Phys. Lett. {\bf B565},
  107  (2003) [arXiv:nucl-th/0305059].

\bibitem{Sinyukov:2004am}
Y.~M. Sinyukov and I.~A. Karpenko, Acta Phys. Hung. {\bf A25},  141  (2006)
  [arXiv:nucl-th/0506002].

\bibitem{Csorgo:2006ax}
T. Cs\"org\H{o}, M.~I. Nagy, and M. Csan\'ad, Phys. Lett. {\bf B663},  306
  (2008) [arXiv:nucl-th/0605070].

\bibitem{Pratt:2004zq}
S. Pratt and S. Pal, Nucl. Phys. {\bf A749},  268  (2005)
  [arXiv:nucl-th/0409038].

\bibitem{Broniowski:2008vp}
W. Broniowski, M. Chojnacki, W. Florkowski, and A. Kisiel,  [arXiv:0801.4361].

\bibitem{Amelin:2007ic}
N.~S. Amelin {\it et~al.}, Phys. Rev. {\bf C77},  014903  (2008)
  [arXiv:0711.0835].

\bibitem{Humanic:2005ye}
T.~J. Humanic, Int. J. Mod. Phys. {\bf E15},  197  (2006)
  [arXiv:nucl-th/0510049].

\bibitem{Tomasik:2005ny}
B. Tom\'a\v{s}ik, AIP Conf. Proc. {\bf 828},  464  (2006)
  [arXiv:nucl-th/0509100].

\bibitem{Csorgo:2005gd}
T. Cs\"org\H{o}, J. Phys. Conf. Ser. {\bf 50},  259  (2006)
  [arXiv:nucl-th/0505019].

\bibitem{Lisa:2005dd}
M.~A. Lisa, S. Pratt, R. Soltz, and U. Wiedemann, Ann. Rev. Nucl. Part. Sci.
  {\bf 55},  357  (2005) [arXiv:nucl-ex/0505014].

\bibitem{Csorgo:2003ry}
T. Cs\"org\H{o}, L.~P. Csernai, Y. Hama, and T. Kodama, Heavy Ion Phys. {\bf
  A21},  73  (2004) [arXiv:nucl-th/0306004].

\bibitem{Csorgo:1995bi}
T. Cs\"org\H{o} and B. L\"orstad, Phys. Rev. {\bf C54},  1390  (1996)
  [arXiv:hep-ph/9509213].

\bibitem{Csanad:2003qa}
M. Csan\'ad, T. Cs\"org\H{o}, and B. L\"orstad, Nucl. Phys. {\bf A742},  80
  (2004) [arXiv:nucl-th/0310040].

\bibitem{Csanad:2003sz}
M. Csan\'ad, T. Cs\"org\H{o}, B. L\"orstad, and A. Ster, Acta Phys. Polon. {\bf
  B35},  191  (2004) [arXiv:nucl-th/0311102].

\bibitem{Csorgo:2004id}
T. Cs\"org\H{o}, M. Csan\'ad, B. L\"orstad, and A. Ster, Acta Phys. Hung. {\bf
  A24},  139  (2005) [arXiv:hep-ph/0406042].

\bibitem{Ster:1999ib}
A. Ster, T. Cs\"org\H{o}, and B. L\"orstad, Nucl. Phys. {\bf A661},  419
  (1999) [arXiv:hep-ph/9907338].

\bibitem{Csorgo:1999sj}
T. Cs\"org\H{o}, Heavy Ion Phys. {\bf 15},  1  (2002) [arXiv:hep-ph/0001233].

\bibitem{Agababyan:1997wd}
N.~M. Agababyan {\it et~al.}, Phys. Lett. {\bf B422},  359  (1998)
  [arXiv:hep-ex/9711009].

\bibitem{Csorgo:1994in}
T. Cs\"org\H{o}, B. L\"orstad, and J. Zim\'anyi, Z. Phys. {\bf C71},  491
  (1996) [arXiv:hep-ph/9411307].

\bibitem{Csorgo:2001ru}
T. Cs\"org\H{o}, Acta Phys. Polon. {\bf B37},  483  (2006)
  [arXiv:hep-ph/0111139].

\bibitem{Csorgo:1998yk}
T. Cs\"org\H{o}, Central Eur. J. Phys. {\bf 2},  556  (2004)
  [arXiv:nucl-th/9809011].

\bibitem{Csorgo:2002kt}
T. Cs\"org\H{o} and J. Zim\'anyi, Heavy Ion Phys. {\bf 17},  281  (2003)
  [arXiv:nucl-th/0206051].

\bibitem{Akkelin:2000ex}
S.~V. Akkelin {\it et~al.}, Phys. Lett. {\bf B505},  64  (2001)
  [arXiv:hep-ph/0012127].

\bibitem{Csorgo:1995vf}
T. Cs\"org\H{o} and B. L\"orstad, Nucl. Phys. {\bf A590},  465c  (1995)
  [arXiv:hep-ph/9503494].

\bibitem{Csanad:2005gv}
M. Csan\'ad {\it et~al.},  [arXiv:nucl-th/0512078].

\bibitem{Adare:2006ti}
A. Adare {\it et~al.}, Phys. Rev. Lett. {\bf 98},  162301  (2007)
  [arXiv:nucl-ex/0608033].

\bibitem{Afanasiev:2007tv}
S. Afanasiev {\it et~al.}, Phys. Rev. Lett. {\bf 99},  052301  (2007)
  [arXiv:nucl-ex/0703024].

\bibitem{Tomasik:2004bn}
B. Tom\'a\v{s}ik, Acta Phys. Polon. {\bf B36},  2087  (2005)
  [arXiv:nucl-th/0409074].

\bibitem{Pratt:1986cc}
S. Pratt, Phys. Rev. {\bf D33},  1314  (1986).

\bibitem{Heinz:2002au}
U.~W. Heinz, A. Hummel, M.~A. Lisa, and U.~A. Wiedemann, Phys. Rev. {\bf C66},
  044903  (2002) [arXiv:nucl-th/0207003].

\bibitem{Tomasik:2002rx}
B. Tom\'a\v{s}ik and U.~A. Wiedemann,  in {\em Quark Gluon Plasma 3},  edited
  by R.~C. Hwa and X.-N. Wang (World Scientific, Singapore, 2004), pp.\
  715--777.
 [arXiv:hep-ph/0210250]
\bibitem{Adler:2003cb}
S.~S. Adler {\it et~al.}, Phys. Rev. {\bf C69},  034909  (2004)
  [arXiv:nucl-ex/0307022].

\bibitem{Adams:2004bi}
J. Adams {\it et~al.}, Phys. Rev. {\bf C72},  014904  (2005)
  [arXiv:nucl-ex/0409033].

\bibitem{Adams:2003ra}
J. Adams {\it et~al.}, Phys. Rev. Lett. {\bf 93},  012301  (2004)
  [arXiv:nucl-ex/0312009].

\end{thebibliography}
\end{document}